\documentclass[iop,apj]{emulateapj}

\usepackage{apjfonts}
\usepackage{natbib}
\usepackage{amssymb}
\usepackage{enumitem}
\usepackage[breaklinks,colorlinks,citecolor=blue,linkcolor=magenta]{hyperref}

\newcommand{\magiicat}{\hbox{{\rm MAG}{\sc ii}CAT}}
\newcommand{\MgIIdblt}{{\rm Mg}\kern 0.1em{\sc ii}~$\lambda\lambda 2796, 2803$}
\newcommand{\MgII}{\hbox{{\rm Mg}\kern 0.1em{\sc ii}}}
\newcommand{\SiII}{\hbox{{\rm Si}\kern 0.1em{\sc ii}}}
\newcommand{\HI}{\hbox{{\rm H}\kern 0.1em{\sc i}}}
\newcommand{\kms}{\hbox{km~s$^{-1}$}}
\newcommand{\etal}{et~al.}
\newcommand{\vfifty}{\hbox{$\Delta v(50)$}}
\newcommand{\vninety}{\hbox{$\Delta v(90)$}}
\newcommand{\face}{\hbox{$i<57^{\circ}$}}
\newcommand{\edge}{\hbox{$i \geq 57^{\circ}$}}
\newcommand{\minor}{\hbox{$\Phi \geq 45^{\circ}$}}
\newcommand{\major}{\hbox{$\Phi<45^{\circ}$}}
\newcommand{\blue}{\hbox{$B-K<1.4$}}
\newcommand{\red}{\hbox{$B-K\geq1.4$}}

\shorttitle{\sc ~{\magiicat} Kinematics: Orientations}
\shortauthors{\sc Nielsen {\etal}}
\slugcomment{Accepted to ApJ, September 12, 2015}

\begin{document}

\title{~{\magiicat} V. Orientation of Outflows and Accretion Determine
  the Kinematics and Column Densities of the Circumgalactic Medium}

\author{
Nikole M. Nielsen$^{1,2}$,
Christopher W. Churchill$^2$,
Glenn G. Kacprzak$^1$,
Michael T. Murphy$^1$,
and
Jessica L. Evans$^2$
}

\affil{$^1$ Centre for Astrophysics and Supercomputing, Swinburne
  University of Technology, Hawthorn, Victoria 3122, Australia; nikolenielsen@swin.edu.au\\
  $^2$ Department of Astronomy, New Mexico State University, Las
  Cruces, NM 88003, USA}

\begin{abstract}

We investigate the dependence of gas kinematics and column densities
in the {\MgII}-absorbing circumgalactic medium on galaxy color,
azimuthal angle, and inclination to trace baryon cycle processes. Our
sample of 30 foreground isolated galaxies at $0.3<z_{\rm gal}<1.0$,
imaged with the {\it Hubble Space Telescope}, are probed by background
quasars within a projected distance of $20<D<110$~kpc. From the
high-resolution ($\Delta v\simeq 6.6$~{\kms}) quasar spectra, we
quantify the extent of the absorber velocity structure with
pixel-velocity two-point correlation functions. Absorbers with the
largest velocity dispersions are associated with blue, face-on
({\face}) galaxies probed along the projected minor axis ({\minor}),
while those with the smallest velocity dispersions belong to red,
face-on galaxies along the minor axis. The velocity structure is
similar for edge-on ({\edge}) galaxies regardless of galaxy color or
azimuthal angle, for red galaxies with azimuthal angle, and for blue
and red galaxies probed along the projected major axis ({\major}). The
cloud column densities for face-on galaxies and red galaxies are
smaller than for edge-on galaxies and blue galaxies,
respectively. These results are consistent with biconical outflows
along the minor axis for star-forming galaxies and accreting and/or
rotating gas, which is most easily observed in edge-on galaxies probed
along the major axis. Gas entrained in outflows may be fragmented with
large velocity dispersions, while gas accreting onto or rotating
around galaxies may be more coherent due to large path lengths and
smaller velocity dispersions. Quiescent galaxies may exhibit
little-to-no outflows along the minor axis, while accretion/rotation
may exist along the major axis.

\end{abstract}


\section{Introduction}
\label{sec:intro}

The circumgalactic medium (CGM) is generally defined as the bound
gaseous halo surrounding galaxies which extends out to a few hundred
kiloparsecs \citep[e.g.,][]{ggk08, steidel10, tumlinson11, rudie12,
  kacprzak13, magiicat2, magiicat1, tumlinson13, werk13}. This region
has increasingly been found to host some of the most important
mechanisms involved in galaxy evolution through the baryon cycle,
including gas accretion via the intergalactic medium and/or recycled
accretion via the galactic fountain mechanism, galactic-scale
outflowing winds, and merging satellite galaxies. The diffuse,
multiphase nature of the CGM lends itself to study by way of
absorption lines found in bright, background objects such as quasars
or galaxies, or even in the spectrum of the host galaxy itself (i.e.,
the ``down-the-barrel'' approach).

Most work examining the low-ionization, cool ($T\sim10^4$~K) component
of the CGM has been focused on {\MgIIdblt} doublet absorption in
background quasar spectra as it is easily observed in the optical at
redshifts $0.1<z<2.5$ \citep[e.g.,][]{bb91, sdp94, guillemin97,
  steidel97, cwc-china, bc09, kcems11, lan14}, and for a range of
{\HI} column densities \citep[$16 \leq \log N({\HI}) \leq 22$,
  e.g.,][]{bs86, ss92, archiveI, rao00, weakII}. The {\MgII} ion is
well-known to be a tracer of the steps involved in the baryon cycle
including accretion, rotating material merging onto the galaxy, and
outflows. {\MgII} absorbing gas may also be associated with merging
satellites which are in the process of being tidally stripped and/or
have ongoing star formation driven outflows.


Accreting gas in the form of filaments from the cosmic web and/or
recycled accretion from past outflows has been found to lie near the
plane of the galaxy disk both in observations
\citep[e.g.,][]{steidel02, ggk-sims, martin12, rubin-accretion,
  bouche13} and in simulations \citep[e.g.,][]{stewart11,
  ford14}. Simulations have also shown that this material forms an
extended (out to $\sim 0.3R_{\rm vir}$), warped disk that co-rotates
with the galaxy when viewed in edge-on orientations
\citep[e.g.,][]{stewart11, danovich12, danovich14}. Direct
observations of infalling material have been few due to the small
covering fraction of the accreting material \citep[at least $6\%$;
  e.g.,][]{martin12, rubin-accretion} and because outflows dominate
the absorption profile, though the spectra used in these works have
low resolution with $\Delta v>150~{\kms}$, which may contribute to the
low detection rate. Nonetheless, accretion and rotation signatures may
include velocities that are bound to the host galaxy, but are greater
than or comparable to projected rotational velocities of the disk for
edge-on galaxies probed along the projected major axis. The velocities
of these absorbers likely lie to one side of the galaxy's systemic
velocity \citep{steidel02, ggk-sims, stewart11, bouche13}, especially
when viewing galaxies in nearly edge-on inclinations.


Based on low-resolution ($\Delta v>150$~{\kms}) spectra, outflows from
galactic-scale winds due to star formation and/or supernovae feedback
are often invoked to explain the presence of {\MgII} absorption
\citep[e.g.,][]{rubin-winds, bouche12, martin12, bordoloi14-model,
  bordoloi14, rubin-winds14, kacprzak14}. A down-the-barrel approach
was used in \citet{weiner09} and \citet{rubin-winds}, and these
authors show that, in stacked galaxy spectra, outflows are commonly
observed in face-on orientations due to the observed blueshift of
material being pushed out of the host galaxy in the direction of the
observer. Using both a down-the-barrel sightline and a transverse
sightline through a single galaxy, \citet{kacprzak14} find that the
sightlines are kinematically coupled and associated with outflowing
material despite the large projected distance (58~kpc) between the
two. In this case, the transverse quasar sightline is aligned with the
galaxy projected minor axis. Finally, by modeling galactic winds to
reproduce the absorption found in seven transverse sightlines, most of
which are aligned with the galaxy projected minor axis,
\citet{bouche12} were able to estimate properties of the winds such as
wind speeds. All of this work is consistent with the picture of
biconical, polar outflows whose signatures include broad, complex
absorption profiles spanning hundreds of {\kms} that cannot be
explained by rotation \citep{outflowsreview}. Such a biconical outflow
model is also applicable to our own Milky Way Galaxy \citep{fox15}.


Merging satellite galaxies may also be a source of {\MgII} absorption
in the CGM. For $z>0.3$ the low luminosities of satellite galaxies
makes it difficult to directly observe such satellites. In fact,
\citet{martin12} found a redshifted absorber with respect to the
targeted galaxy and associated it with a satellite that was clearly
detected in their galaxy images, but this was the only case out of
their sample of over 200 galaxies, indicating that satellites as the
source of absorption is likely rare ($<1\%$ probability). Examining
the azimuthal angle dependence of satellites around host galaxies,
\citet{yang06} found that satellites within $D=700$~kpc tend to be
isotropically distributed around blue galaxies, but are preferentially
located along the major axis of red galaxies. This result has been
found with a variety of data sets, which are summarized in
\citet{yang06}, as well as in simulations \citep[most recently
  by][]{dong14}.

\begin{deluxetable*}{lrccll}
\tablecolumns{6} 
\tablewidth{0pt} 
\tablecaption{TPCF Velocity Measurements \label{tab:v50}}
\tablehead{
\colhead{Galaxy Sample} &
\colhead{\# Gals} &
\colhead{Cut} &
\colhead{Cut} &
\colhead{$\Delta v(50)$\tablenotemark{a}} &
\colhead{$\Delta v(90)$\tablenotemark{a}} 
}
\startdata
\cutinhead{Figure~\ref{fig:PAincl}}\\[-3pt]
Face-on & 17 & {\face} & $\cdots$ & \phn$ 69_{- 9}^{+ 8}$ & $200_{-21}^{+15}$\\[3pt] 
Edge-on & 13 & {\edge} & $\cdots$ & \phn$ 64_{- 6}^{+ 4}$ & $149_{-14}^{+10}$\\[10pt] 

Major Axis & 15 & {\major} & $\cdots$ & \phn$ 55_{-13}^{+ 7}$ & $146_{-32}^{+13}$\\[3pt]
Minor Axis & 15 & {\minor} & $\cdots$ & \phn$ 73_{- 7}^{+ 5}$ & $192_{-21}^{+15}$\\[10pt]

Face-on -- Major Axis & 10 & {\face} & {\major} & \phn$ 45_{- 6}^{+ 6}$ & $124_{-22}^{+20}$\\[3pt]
Face-on -- Minor Axis &  7 & {\face} & {\minor} & \phn$ 84_{-13}^{+ 8}$ & $220_{-25}^{+13}$\\[3pt]
Edge-on -- Major Axis &  5 & {\edge} & {\major} & \phn$ 73_{-20}^{+10}$ & $171_{-45}^{+23}$\\[3pt]
Edge-on -- Minor Axis &  8 & {\edge} & {\minor} & \phn$ 60_{- 5}^{+ 4}$ & $141_{-11}^{+ 8}$\\[3pt]

\cutinhead{Figure~\ref{fig:BKPA}}\\[-3pt]
Blue -- Major Axis &  5 & {\blue} & {\major} & \phn$ 61_{-11}^{+ 9}$ & $162_{-23}^{+20}$\\[3pt]
Blue -- Minor Axis & 10 & {\blue} & {\minor} & \phn$ 81_{- 9}^{+ 8}$ & $209_{-23}^{+16}$\\[3pt]
Red -- Major Axis  & 10 & {\red}  & {\major} & \phn$ 51_{-13}^{+10}$ & $135_{-36}^{+20}$\\[3pt]
Red -- Minor Axis  &  5 & {\red}  & {\minor} & \phn$ 52_{- 4}^{+ 1}$ & $123_{- 9}^{+ 2}$\\[3pt]

\cutinhead{Figure~\ref{fig:BKincl}}\\[-3pt]
Blue -- Face-on &  8 & {\blue} & {\face} & \phn$ 88_{-10}^{+ 8}$ &     $227_{-19}^{+12}$\\[3pt]
Blue -- Edge-on &  7 & {\blue} & {\edge} & \phn$ 61_{- 8}^{+ 4}$ &     $143_{-19}^{+ 9}$\\[3pt]
Red -- Face-on  &  9 & {\red}  & {\face} & \phn$ 40_{- 4}^{+ 3}$ & \phn$ 97_{- 9}^{+ 6}$\\[3pt]
Red -- Edge-on  &  6 & {\red}  & {\edge} & \phn$ 62_{-14}^{+ 7}$ &     $147_{-33}^{+15}$\\[-5pt]
\enddata
\tablenotetext{a}{{\kms}}
\end{deluxetable*}

Detailed absorber kinematics for a large absorber--galaxy sample,
which are required to provide insight into the motions of the gas
involved in the baryon cycle and constrain simulations, have thus far
remained elusive due to the need to stack low-resolution spectra. Even
without needing to stack the spectra, much of the velocity structure
is washed out by the low resolution. Most of the previous work
examining {\MgII} absorbers associated with baryon cycle processes
have reported that the {\MgII} equivalent width depends strongly on
inclination and/or azimuthal angle \citep[e.g.,][]{bordoloi11,
  kcems11, bouche12, kcn12, bordoloi14-model, bordoloi14}, however
equivalent width does not provide a detailed picture of the gas
kinematics. This is especially true since equivalent width depends on
the velocity spread of absorption {\it and} the column densities,
which in turn depend on the line-of-sight geometry, metallicity, and
ionization conditions of the absorbers. We therefore make use of a
subsample of galaxies in the {\MgII} Absorber--Galaxy Catalog
\citep[{\magiicat};][]{magiicat2, magiicat1, cwc-masses2} for which we
have high-resolution (${\rm R} \sim 45,000$, $\Delta v \simeq
6.6$~{\kms}) background quasar spectra, in addition to the colors and
orientations of the galaxies themselves. With these line-of-sight
data, we can study the detailed kinematics of {\MgII} absorbers and
examine if the enhanced detection rates and larger equivalent widths
for particular orientations are due to velocity spreads, column
densities, or some combination of both.

In a previous work, \citet{kcn12} examined the equivalent widths of
{\MgII} absorbers with galaxy orientation for a subset of
absorber--galaxy pairs from {\magiicat} \citep{magiicat2, magiicat1}
by modeling the galaxies and measuring the azimuthal angles at which
background quasars probe the CGM. They found that {\MgII} absorption
prefers to be located along the major and minor axes of blue galaxies,
while there is no such preference for absorbers located around red
galaxies \citep[a result confirmed with a larger, statistical sample
  by][]{lan14}, or for sightlines in which no absorption is detected
(nonabsorbers) regardless of galaxy color. \citet{kcn12} also found a
tendency for {\MgII} equivalent widths to be larger along the minor
axis than the major axis, possibly an indication of probing more
enriched outflowing material along the minor axis.

In this paper, we expand upon our previous work and examine the
kinematics of absorbing gas as a function of galaxy color and
inclination as well as the azimuthal angle at which the galaxy is
probed by using {\MgII} absorber pixel-velocity two-point correlation
functions \citep[TPCFs, described in detail by][]{nielsen-TPCF1} and
cloud column densities. This paper is organized as
follows. Section~\ref{sec:methods} describes the sample, our methods
for analyzing the spectra and data, and how we calculate the
TPCFs. Section~\ref{sec:results} details how the absorber kinematics
(TPCFs) and cloud column densities differ for galaxies of various
colors, azimuthal angles, and inclinations. In
Section~\ref{sec:discussion} we place our kinematics results in the
context of the baryon cycle and discuss the implications. Finally, we
summarize and conclude our findings in
Section~\ref{sec:conclusions}. We adopt a $\Lambda$CDM cosmology
($H_{0}=70$~{\kms}~Mpc$^{-1}$, $\Omega_{M}=0.3$,
$\Omega_{\Lambda}=0.7$) throughout this paper.

\section{Sample and Data Analysis}
\label{sec:methods}

\subsection{Data}
\label{sec:methodsdata}

We use a sample of 30 spectroscopically confirmed ($0.3<z_{\rm
  gal}<1.0$) {\MgII} absorption-selected galaxies from the {\MgII}
Absorber--Galaxy Catalog \citep[{\magiicat;}][]{magiicat2,
  magiicat1}. The galaxies are isolated to the limits of the data,
where isolation is defined as having no spectroscopically identified
neighbor within a projected distance of 100~kpc or a line-of-sight
velocity of 500~{\kms}. Each galaxy has been imaged with WFPC2/{\it
  HST} in the F702W band and we have a rest-frame $B-K$ color for each
galaxy which was determined as described in \citet{magiicat1}.

\begin{figure}[ht]
\centering
\includegraphics[scale=1]{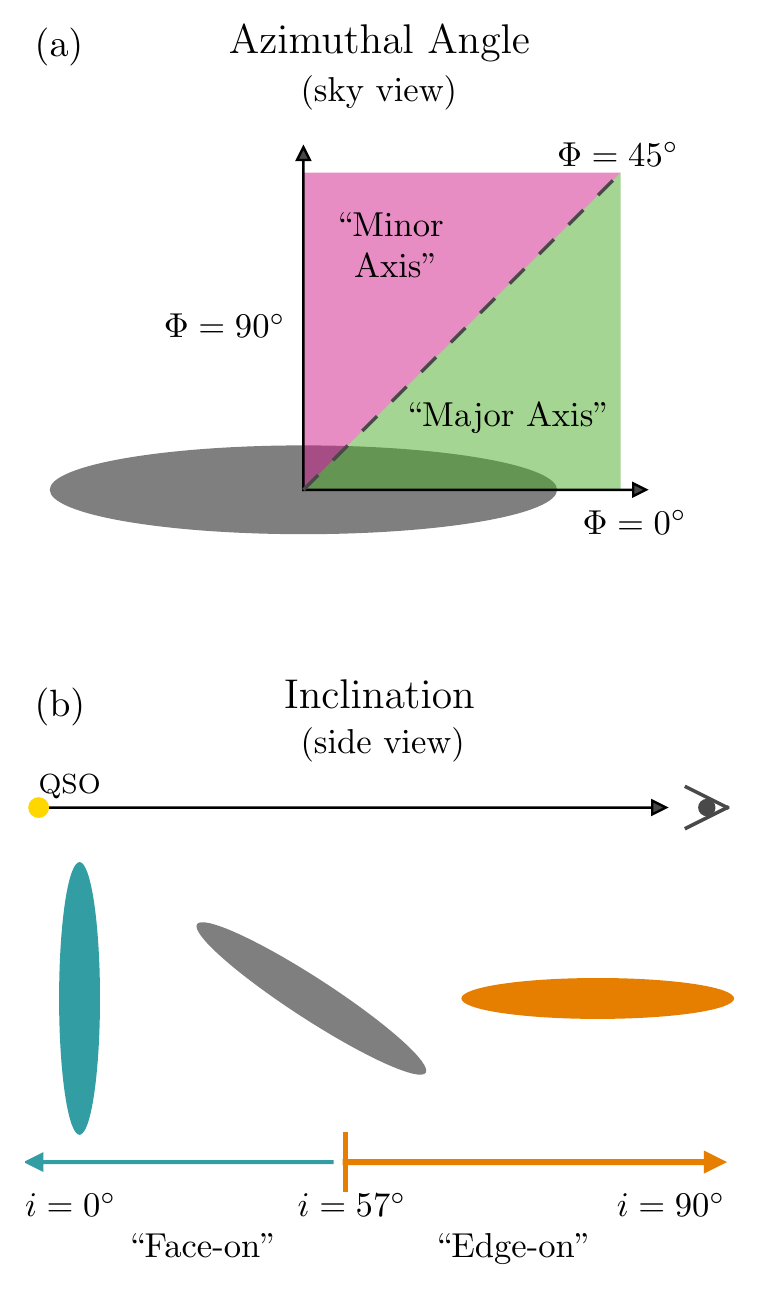}
\caption[]{Diagram demonstrating galaxy azimuthal angle and
  inclination subsamples. Panel (a) presents an on-the-sky view of the
  azimuthal angle around an inclined galaxy (gray ellipse). An
  azimuthal angle of $\Phi=0^{\circ}$ is defined as the projected
  galaxy major axis, while $\Phi=90^{\circ}$ is the projected galaxy
  minor axis. Galaxies which are probed by quasar sightlines at
  $0^{\circ}\leq \Phi < 45^{\circ}$ are included in the ``major axis''
  subsample (green shaded region), while those probed at $45^{\circ}
  \leq \Phi \leq 90^{\circ}$ are included in the ``minor axis''
  subsample (pink shaded region). Colors in this panel correspond to
  the colors in Figures~\ref{fig:PAincl}, \ref{fig:BKPA}, and
  \ref{fig:InclNvsV}. Panel (b) presents a side view of three galaxies
  with different inclinations whose CGM is probed by a background
  quasar. We define ``face-on'' galaxies as those with $0^{\circ}\leq
  i <57^{\circ}$ (blue ellipse and arrow) while ``edge-on'' galaxies
  are those with $57^{\circ} \leq i \leq 90^{\circ}$ (orange ellipse
  and arrow, including the gray ellipse). An inclination of
  $i=57^{\circ}$ is the value for which we cut our sample and
  corresponds to the minimum inclination for an ``edge-on'' galaxy
  (gray ellipse). Colors and arrow line widths in this panel
  correspond to the colors in Figures~\ref{fig:PAincl},
  \ref{fig:BKincl}, and \ref{fig:PANvsV}.}
\label{fig:diagram}
\end{figure}

All galaxies were modeled using GIM2D \citep{simard02} to obtain
quantified morphological parameters, inclinations, and position
angles. Full details of the method used and the morphological
properties of most galaxies in our sample are presented in
\citet{kcems11}. We define inclinations of $i=0^{\circ}$ as face-on
and $i=90^{\circ}$ as edge-on. We convert position angles to an
``azimuthal angle'' which describes where a background quasar
sightline is located with respect to the projected major axis of the
galaxy. An azimuthal angle of $\Phi=0^{\circ}$ is defined as having
the quasar line of sight along the projected galaxy major axis and
$\Phi=90^{\circ}$ as having the sightline along the projected minor
axis. Figure~\ref{fig:diagram} illustrates our galaxy orientation
definitions with azimuthal angle in panel (a) and inclination in panel
(b).

The CGM of each galaxy is probed by a nearby (projected on the sky
distance of $20<D<110$~kpc) background quasar for which we have a
high-resolution HIRES/Keck or UVES/VLT spectrum. We refer the reader
to \citet{cwc-thesis}, \citet{cv01}, \citet{cvc03},
\citet{evans-thesis}, and \citet{kcems11} for the spectra, full
details of the reduction and analysis, and the detection and Voigt
profile (VP) fitting of {\MgIIdblt} absorption. We measure {\MgII}
equivalent widths and, from the VP fitting, we obtain VP component
(cloud) column densities, velocities, and Doppler $b$
parameters. Velocity zero points are defined as the median velocity of
the apparent optical depth distribution of {\MgII} absorption
\citep{cwc-thesis}.

For our TPCF analysis, we use only the velocities of pixels that reside
in regions where absorption is formally detected. These regions are
defined as ``kinematic subsystems'' in \citet{cv01} and their velocity
bounds are determined by searching the spectra to either side of the
subsystem centroids for the point at which the significance in the per
pixel equivalent width falls below $1\sigma$. We enforce a sensitivity
cut to these regions of $W_r(2796)\geq 0.04$~{\AA} to account for
differences in the quality of our spectra \citep{nielsen-TPCF1}. We
are $\sim 95\%$ complete to this detection sensitivity within $\pm
800$~{\kms} for all absorbers in our sample. This sensitivity cut is
used to ensure that we can detect absorption uniformly throughout our
sample. Thus, the results in this paper apply to {\MgII} absorption
with $W_r(2796)\geq 0.04$~{\AA}, and therefore samples only those
regions of the CGM for which the temperature, metallicity, ionization,
and line-of-sight conditions give rise to gas detected with {\MgII}
for our sample.

We slice the sample into several subsamples based on galaxy rest-frame
$B-K$ color, azimuthal angle, $\Phi$, and inclination, $i$. Galaxy
color and azimuthal angle cuts were determined by calculating the
median values for the sample we present here \citep[similar to the
  method used in][]{magiicat2, magiicat1}. Our inclination cut was
defined by the average inclination of galaxies in the universe,
$i=57.3^{\circ}$ \citep[for a derivation, see the appendix
  of][]{law09-avgi}. We tabulate the characteristics of each subsample
in Table~\ref{tab:v50} including the subsample names, number of
galaxies in each subsample, and the median value(s) by which the
sample was cut. The $B-K$ color cut we use to separate our sample into
``blue'' and ``red'' galaxies is strictly to indicate whether the
galaxy is more likely to be star-forming (blue) or passive (red),
rather than indicating morphological types such as early-type or
late-type galaxies. Figure~\ref{fig:diagram} illustrates our subsample
cuts for azimuthal angle (panel (a)) and inclination (panel (b)).

We caution that, in our sample, a weak trend exists between $B-K$ and
halo mass, $\log (M_{\rm h}/M_{\odot})$, (Kendall $\tau$ rank
correlation test, $2.1\sigma$) where redder galaxies tend to be more
massive. We find that, with the exception of eight galaxies, blue
galaxies tend to be low mass while red galaxies tend to be high mass
\citep[see][for details]{nielsen-TPCF1}. Therefore, the color
dependencies in our results are more accurately color--mass
dependencies.

To rule out the possibility that any differences in our results are
due to biased distributions in azimuthal angle and inclination, we ran
a one-dimensional Kolmogorov--Smirnov (KS) test for both orientation
measures. We find that the azimuthal angles and inclinations of
galaxies in our sample are consistent with unbiased samples at the
$0.6\sigma$ and $2.3\sigma$ levels, respectively. Additionally,
rank-correlation tests between $\Phi$ or $i$ and galaxy properties
such as rest-frame $B-K$ color show no correlations. Therefore, any
differences we see in our results are likely not due to underlying
sample biases.

\subsection{Pixel-Velocity Two-Point Correlation Functions}
\label{sec:methodsTPCF}

Throughout this paper we examine absorber pixel-velocity two-point
correlation functions (TPCFs), which are a measure of the {\it
  internal} absorber velocity dispersion. We remind the reader that
the absorber--galaxy sample presented here is an absorption-selected
sample with an equivalent width detection threshold of $W_r(2796)\geq
0.4$~{\AA} (see Section~\ref{sec:methodsdata}). 

To construct the TPCF, we first define a subsample of galaxies, e.g.,
blue galaxies, and examine the associated absorbers. We obtain the
velocities of all pixels that reside only in spectral regions where
{\MgII}~$\lambda 2796$ absorption is formally detected using the
detection methods of \citet{cv01}, who refer to these regions as
``kinematic subsystems.'' After pooling all of the absorbing pixels
from every line-of-sight in the galaxy subsample together, we then
calculate the velocity differences of each possible pair of pixels and
take their absolute value to get $\Delta v_{\rm pixel}$. We bin up
these pixel-velocity separations and normalize each bin by the total
number of pixel pairs in the subsample for comparison between
different galaxy subsamples. Thus, the TPCF is a probability
distribution function. We use a bin size of 10~{\kms}, which
corresponds to roughly one resolution element of both the HIRES/Keck
and UVES/VLT spectrographs (three pixels per resolution element, with
a resolution of $\sim 6.6$~\kms).

To obtain the uncertainties on the TPCF, we conduct a bootstrap
analysis for 100 realizations, where the maximum number of
realizations allowed for a sample size of five (our smallest sample)
with replacement is 126. More realizations than this begin repeating
permutations too often. We calculate $1\sigma$ standard deviations
from the mean of the bootstrap realizations in each bin.

We quantitatively characterize the TPCFs by measuring the velocity
separations within which 50\% and 90\% of the area under the TPCF
distribution is located, {\vfifty} and {\vninety},
respectively. Uncertainties on these values are obtained from the
bootstrap analysis and represent $1\sigma$ deviations from the mean of
the bootstrap realizations. The values for {\vfifty} and {\vninety}
are tabulated in Table~\ref{tab:v50} for each subsample.

\begin{figure*}[ht]
\centering
\includegraphics[scale=0.7]{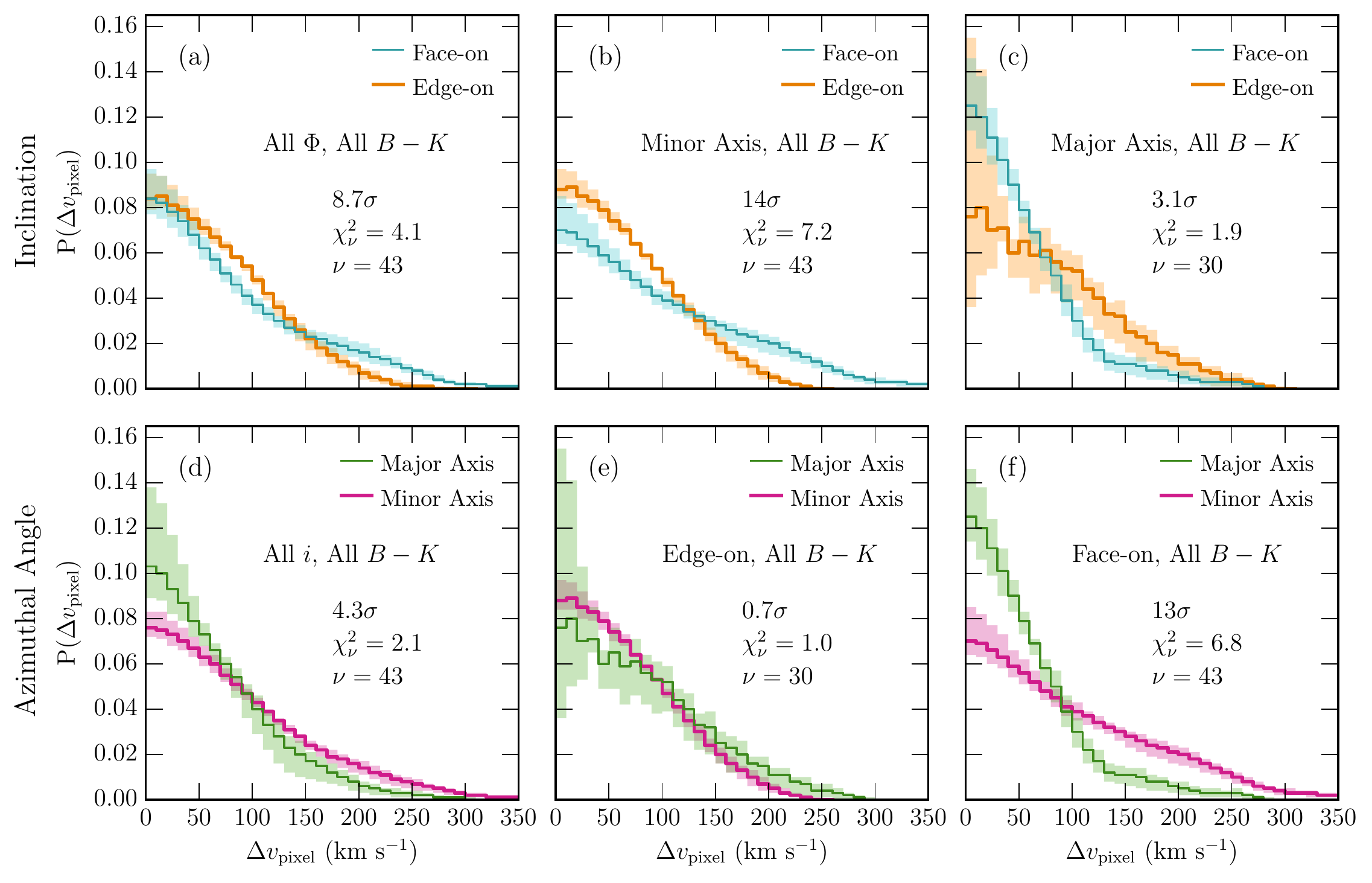}
\caption[]{Pixel-velocity two-point correlation functions examining
  how the spread in the pixel velocities of absorbers differs when
  probing galaxies of different inclinations and azimuthal angles (for
  all galaxy colors, $B-K$). Panels (a), (b), and (c) compare TPCFs of
  face-on ({\face}) and edge-on ({\edge}) galaxies for all $\Phi$,
  {\minor}, and {\major}, respectively. Panels (d), (e), and (f)
  examine how the TPCFs along the major axis ({\major}) and minor axis
  ({\minor}) differ for all $i$, {\edge}, and {\face},
  respectively. In each panel the TPCF is represented as a solid line
  while shaded regions are the $1\sigma$ bootstrap uncertainties. We
  list the significance of a chi-squared test comparing the TPCFs in
  each panel in addition to the reduced chi-squared value,
  $\chi^2_{\nu}$, and the degrees of freedom, $\nu$. We see a broader
  velocity dispersion for face-on galaxies than for edge-on galaxies
  when all azimuthal angles are considered (panel (a)). We also see a
  larger velocity dispersion for galaxies probed along the minor axis
  than along the major axis when all inclinations are considered
  (panel (d)). This latter result becomes highly significant for
  face-on galaxies where the minor axis dispersion is greater than
  that for the major axis (panel (f)). Conversely, there is no
  difference in the velocity dispersion with azimuthal angle for
  edge-on orientations (panel (e)). Thus, the velocity dispersion is
  greatest for face-on galaxies probed along the projected minor
  axis.}
\label{fig:PAincl}
\end{figure*}

\begin{figure*}[ht]
\includegraphics[width=\linewidth]{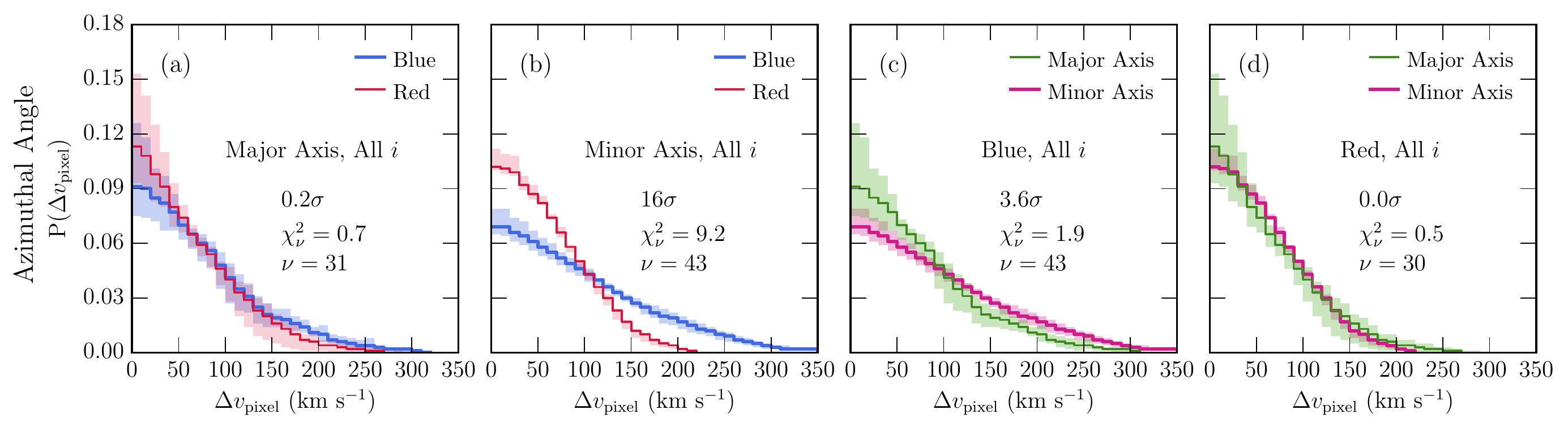}
\caption[]{Pixel-velocity two-point correlation functions for
  subsamples cut by galaxy rest-frame $B-K$ color and azimuthal
  angle. Lines, shading, and the results of a chi-squared test
  comparing subsamples are plotted as in Figure~\ref{fig:PAincl}. For
  all inclinations, blue and red galaxy TPCFs are compared along the
  major axis ({\major}) and minor axis ({\minor}) in panels (a) and
  (b), respectively. Panels (c) and (d) present the TPCFs of blue and
  red galaxies, respectively, at different azimuthal angles. The
  velocity dispersions are all statistically consistent (panels (a)
  and (d)) with the exception of large dispersions for the blue, minor
  axis subsample.}
\label{fig:BKPA}
\end{figure*}

\begin{figure*}[ht]
\includegraphics[width=\linewidth]{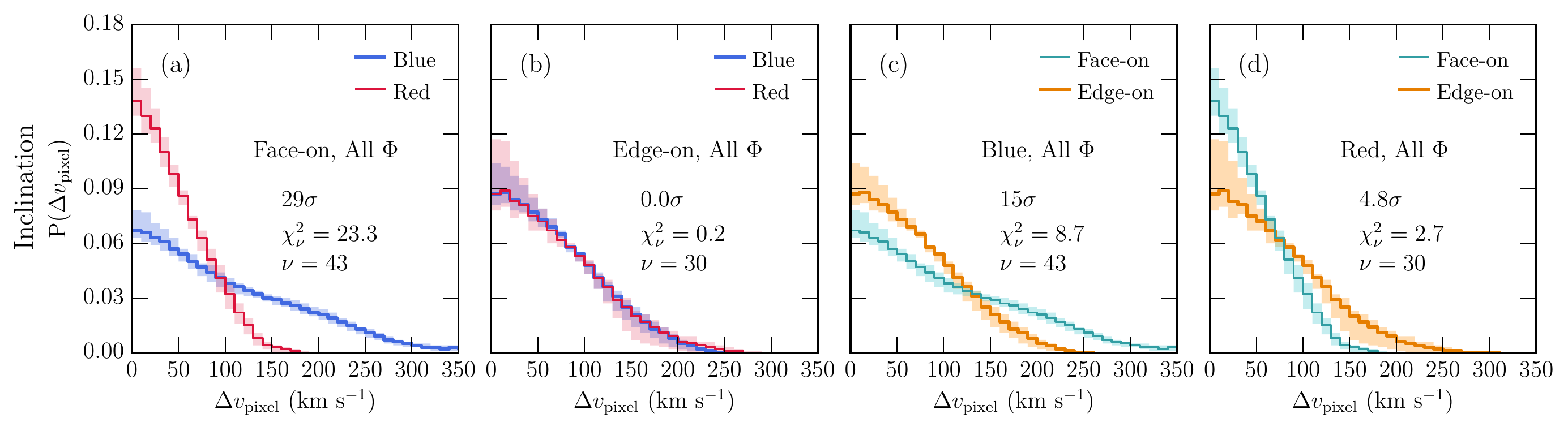}
\caption[]{Pixel-velocity two-point correlation functions for
  subsamples cut by galaxy rest-frame $B-K$ color and
  inclination. Lines, shading, and the results of a chi-squared test
  comparing subsamples are plotted as in Figure~\ref{fig:PAincl}. For
  all azimuthal angles, we compare the TPCFs of blue and red face-on
  ({\face}) galaxies in panel (a) and edge-on ({\edge}) galaxies in
  panel (b). Panels (c) and (d) compare the TPCFs of blue and red
  galaxies, respectively, at different inclinations. We find dramatic
  differences between the absorber velocity dispersions for blue and
  red galaxies for face-on orientations (panel (a)) where blue galaxies
  have much larger velocity dispersions than red galaxies, but we find
  no differences between blue and red for edge-on orientations (panel
  (b)).}
\label{fig:BKincl}
\end{figure*}

\section{Results}
\label{sec:results}

\subsection{TPCFs: Galaxy Inclinations and Azimuthal Angles}

In Figure~\ref{fig:PAincl} we present TPCFs for galaxy subsamples cut
by galaxy azimuthal angle, $\Phi$, and/or inclination, $i$, for all
galaxy colors, $B-K$. Solid lines represent the TPCF for each
subsample while shaded regions indicate the $1\sigma$ uncertainties on
the TPCFs from the bootstrap analysis described in
Section~\ref{sec:methodsTPCF}. Colors for azimuthal angle and
inclination subsamples correspond to those in
Figure~\ref{fig:diagram}(a) and (b), respectively. In the top row we
compare face-on and edge-on galaxy subsamples at all azimuthal angles
(panel (a)), along the projected minor axis (panel (b)), and along the
projected major axis (panel (c)). In the bottom row, we compare
galaxies probed along the major and minor axes for galaxies at all
inclinations (panel (d)), edge-on galaxy subsamples (panel (e)), and
face-on galaxy subsamples (panel (f)). In each panel we present the
results of a chi-squared test comparing the binned data (TPCFs) for
each plotted subsample pair, including the significance, reduced
chi-squared value, $\chi^2_{\nu}$, and degrees of freedom, $\nu$.

In Figure~\ref{fig:PAincl}(a) we examine bivariate trends in the
absorber TPCFs for galaxies with ``face-on'' ({\face}) and ``edge-on''
({\edge}) inclinations for all colors and probed at all azimuthal
angles. Absorbers in face-on galaxies have larger velocity dispersions
than those in edge-on galaxies. A chi-squared test yields the result
that the null hypothesis that the TPCFs are drawn from the same
population can be ruled out at the $8.7\sigma$ level. While the
chi-squared test result is significant, we find that the {\vfifty}
measurements are consistent within uncertainties for face-on and
edge-on galaxies but the value of {\vninety} is larger for face-on
galaxies. Therefore, the difference between these two distributions is
due to the large velocity dispersion tail in face-on galaxies.

We also examine absorption associated with galaxies probed along the
``major axis'' ({\major}) and ``minor axis'' ({\minor}) in
Figure~\ref{fig:PAincl}(d) for all inclinations and colors and find
that absorbers located in galaxies probed along the minor axis have
larger velocity dispersions than those along the major axis
($4.3\sigma$). Both {\vfifty} and {\vninety} are larger for the minor
axis sample than the major axis. For this TPCF pair and the rest in
this paper, the measurements of both {\vfifty} and {\vninety} reflect
the chi-squared test results, i.e., where we find an insignificant
chi-squared value when comparing galaxy subsamples, we also find
values of {\vfifty} and {\vninety} that are consistent within
uncertainties between subsamples.

While we find significant differences in the absorber TPCFs with
inclination or azimuthal angle, both orientation measures are
important for describing the precise location of an absorber around a
galaxy. Therefore, in Figure~\ref{fig:PAincl} we present a
multivariate analysis comparing the absorber TPCFs for face-on and
edge-on galaxies probed along the minor axis (panel (b)) and the major
axis (panel (c)). In both panels we find that the velocity structure
for absorbers probed in face-on galaxies is significantly different
from those in edge-on galaxies. Along the minor (major) axis, the
velocity dispersion of absorbers is greater (smaller) for face-on
galaxies than edge-on galaxies with a significance of $14\sigma$
($3.1\sigma$) in panel (b) (panel (c)).

We also compare absorber TPCFs for galaxies probed along the major and
minor axes for edge-on galaxies (panel (e)) and face-on galaxies
(panel (f)). We find that absorbers located in edge-on galaxies have
similar velocity dispersions regardless of the azimuthal angle at
which the galaxy is probed (panel (e), $0.7\sigma$). On the other
hand, we find a highly significant difference ($13\sigma$) for face-on
galaxies probed along the major and minor axes in
Figure~\ref{fig:PAincl}(f) where the absorber velocity dispersions are
much greater along the minor axis than the major axis. In fact, both
{\vfifty} and {\vninety} for the face-on, minor axis subsample are
roughly twice as large as the face-on, major axis subsample.

\begin{figure*}[ht]
\includegraphics[width=\linewidth]{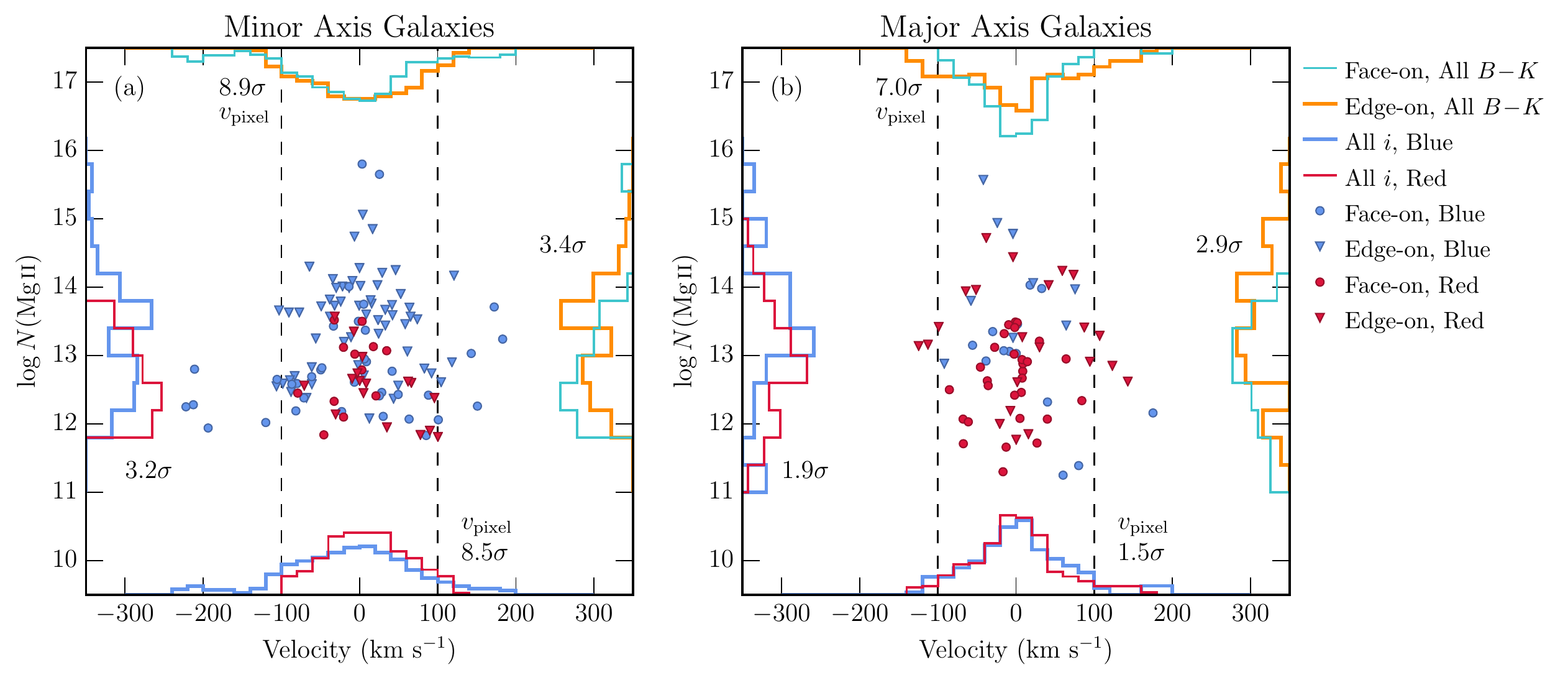}
\caption[]{Cloud (VP component) column densities and velocities, and
  pixel velocities for galaxy subsamples sliced by inclination and
  galaxy color for galaxies probed along (a) the minor axis ({\minor})
  and (b) the major axis ({\major}). Scatter plots show the modeled
  cloud velocities and column densities. Blue points represent blue
  galaxies, red points are red galaxies, circles are face-on galaxies,
  and diamonds are edge-on galaxies. Vertical dashed lines at $\pm
  100$~{\kms} are provided to help guide the reader's eye between
  panels. Histograms on the left and right of each panel show the
  distribution of cloud column densities for subsamples sliced by
  (left) galaxy color, $B-K$, and (right) inclination, $i$. The quoted
  significance near the column density histograms is the result of a
  KS test between plotted subsamples. Top and bottom histograms of
  each panel present the distribution of pixel (not cloud) velocities
  for subsamples cut by (bottom) galaxy color and (top)
  inclination. The significances quoted for the top and bottom
  histograms are for an F-test comparing the variance in the
  distribution of pixel velocities for each pair of plotted
  subsamples. All histograms are normalized by the total number of
  data points in each subsample. In general, we find that clouds with
  larger velocities relative to the optical depth median velocity have
  smaller column densities, while those with larger column densities
  have smaller velocities relative to the optical depth median
  velocity. We find that the cloud column density distributions are
  statistically consistent for galaxies probed along the major axis
  (panel (b)), regardless of whether we compare subsamples sliced by
  galaxy color or inclination. However, we find statistically larger
  column densities for blue and face-on galaxies probed along the
  minor axis (panel (a)).}
\label{fig:PANvsV}
\end{figure*}

\begin{figure*}[ht]
\includegraphics[width=\linewidth]{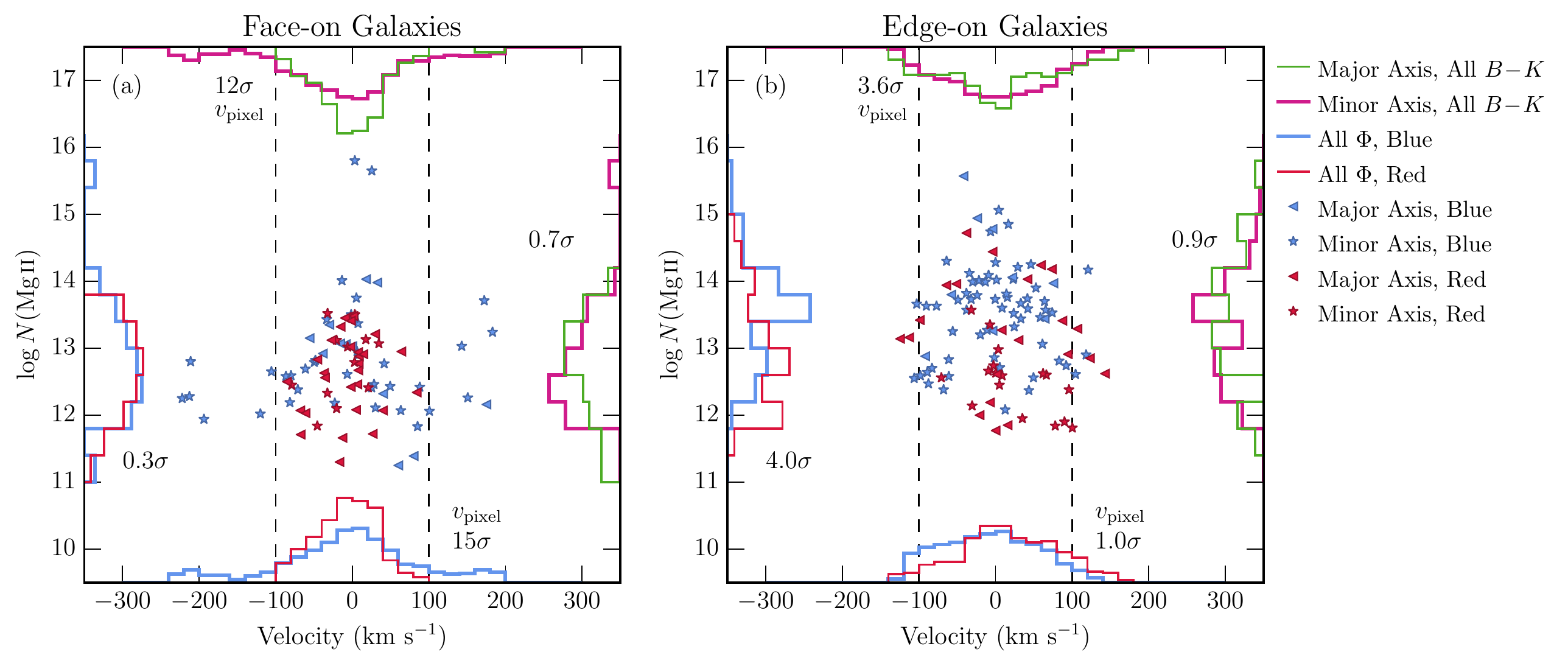}
\caption[]{Cloud column densities and velocities, and pixel velocities
  for (a) face-on galaxies ({\face}) and (b) edge-on galaxies
  ({\edge}) split by galaxy color and azimuthal angle. Blue galaxies
  are represented as blue points, red galaxies as red points, galaxies
  probed along the major axis as triangles, and galaxies probed along
  the minor axis as stars. Vertical dashed lines are plotted to help
  guide the reader's eye between panels. Histograms in each panel
  represent the distributions of cloud column densities split by
  (left) galaxy color, $B-K$, (right) azimuthal angle, $\Phi$, and
  pixel velocities (rather than cloud velocities) split by (bottom)
  galaxy color, and (top) azimuthal angle. The quoted significances
  near the histograms report the results of a KS test between plotted
  subsamples for the cloud column densities (left and right) and the
  results of an F-test comparing the variance in the pixel velocity
  distributions between plotted subsamples for the top and bottom
  histograms. We find that the cloud column densities for absorbers
  around face-on galaxies rarely exceed $\log N({\MgII})=14$ (panel
  (a)), in contrast to edge-on galaxies (panel (b)). For both face-on
  and edge-on galaxies, we see no difference in the column densities
  with regard to the azimuthal angle (right histograms). Lastly, we
  find that the cloud column densities in absorbers around blue
  galaxies with edge-on inclinations are greater than those for
  absorbers around blue, face-on galaxies ($4.7\sigma$, not plotted)
  or red, edge-on galaxies ($4.0\sigma$, panel (b)).}
\label{fig:InclNvsV}
\end{figure*}

\subsection{TPCFs: Galaxy Colors and Orientations}

We also study the differences in absorber TPCFs for galaxies of
various orientations and galaxy color. In Figure~\ref{fig:BKPA}, we
present subsamples cut by azimuthal angle, $\Phi$, and galaxy
rest-frame color, $B-K$, for all galaxy inclinations. We find no
significant difference ($0.2\sigma$) in absorber velocity dispersions
along the major axis in ``blue'' ({\blue}) and ``red'' ({\red})
galaxies (panel (a)) or for absorbers hosted by red galaxies along the
major and minor axes (panel (d)). The only TPCF that is significantly
different from the rest of the subsamples is the blue, minor axis
subsample. This blue, minor axis subsample has a significantly larger
absorber velocity dispersion than either the red, minor axis subsample
(panel (b), $16\sigma$) or the blue, major axis subsample (panel (c),
$3.6\sigma$). While {\vfifty} and {\vninety} are nearly identical for
the red, major axis subsample, the red, minor axis subsample, and the
blue, major axis subsample, the measurements for the blue, minor axis
subsample are larger than the rest.

In Figure~\ref{fig:BKincl}, we examine absorber TPCFs of subsamples
sliced by inclination, $i$, and $B-K$ color. Absorbers located around
edge-on galaxies have similar velocity dispersions (for all $\Phi$)
regardless of galaxy color ($0\sigma$,
Figure~\ref{fig:BKincl}(b)). The values of {\vfifty} and {\vninety}
for edge-on blue and red subsamples are nearly identical and
consistent within uncertainties. On the other hand, absorbers around
face-on galaxies at all azimuthal angles in Figure~\ref{fig:BKincl}(a)
are dramatically different for blue and red galaxies. The absorber
velocity dispersions are much greater ($29\sigma$) for face-on, blue
galaxies than for face-on, red galaxies, which are more highly peaked
at low velocity separations. The {\vfifty} and {\vninety} measurements
are roughly twice as large for the face-on, blue subsample than the
face-on, red subsample. The velocity dispersions of absorbers hosted
by both blue and red galaxies in Figures~\ref{fig:BKincl}(c) and
\ref{fig:BKincl}(d) depend on the observed inclination, but differ in
their trends; blue (red) galaxies have larger velocity dispersions for
face-on (edge-on) inclinations than for edge-on (face-on)
inclinations. This is due to the significant differences in face-on
inclinations (Figure~\ref{fig:BKincl}(a)) between blue and red
galaxies (larger velocity dispersions for blue galaxies than red
galaxies) and the nearly identical TPCFs for edge-on blue and red
galaxies in panel (b).

\subsection{Cloud Column Densities and Velocities}

To obtain further understanding of the CGM material being probed for
each orientation presented in the TPCFs, we plot cloud (VP component)
column densities and velocities in Figure~\ref{fig:PANvsV}. Clouds
plotted in panel (a) are associated with galaxies probed along the
minor axis. In panel (b), clouds are associated with galaxies probed
along the major axis. Scatter plots show the cloud column densities
and velocities from Voigt profile fitting, where vertical dashed lines
on the scatter plots at $\pm 100$~{\kms} are to help guide the eye
between panels. Points are colored by galaxy color, $B-K$, while point
types represent galaxy inclination, $i$. Histograms to the left and
right in each panel present the distribution of cloud column densities
for galaxies sliced by (left) galaxy color and (right) inclination,
where the indicated significance is the result of a KS test between
subsamples for the unbinned data. Finally, the histograms on top and
bottom show the distribution of pixel velocities (rather than cloud
velocities) used to calculate the TPCFs for subsamples cut by galaxy
color (bottom) and inclination (top). The result of an
F-test\footnote{The F-test is designed to measure whether two sample
  distributions have similar or different dispersions; it is sensitive
  to the tails of the distributions.} between the unbinned pixel
velocities of each subsample pair is indicated next to the top and
bottom histograms. In each case, the pixel velocity F-test results
show similar results as the TPCF chi-squared test, i.e., a significant
chi-squared test corresponds to a significant F-test, and vice versa.

In general, we find that higher velocity material has lower column
density values, while higher column density material is only found at
lower velocity. This is consistent with previous works
\citep[e.g.,][]{cvc03}. The highest velocity ($v > |100|$~{\kms}), low
column density clouds are mostly associated with blue, face-on
galaxies probed along the minor axis. There is also a population of $v
\sim |100|$~{\kms} clouds that are associated with edge-on galaxies
probed along the minor axis, though we find a few for major axis
galaxies as well.  All of these clouds at high velocity are found in
absorbers which have higher column density material located near the
velocity zero point.

For minor axis galaxies in Figure~\ref{fig:PANvsV}(a), we find that
blue galaxies (all inclinations) have significantly larger column
densities than red galaxies ($3.2\sigma$) and edge-on galaxies (all
colors) have significantly larger column densities than face-on
galaxies ($3.4\sigma$). For major axis galaxies in
Figure~\ref{fig:PANvsV}(b), we find only a suggestion of larger column
densities for edge-on galaxies than face-on galaxies ($2.9\sigma$, all
colors), though clouds at $\log N({\MgII})>14$ are mainly hosted by
edge-on galaxies. Conversely, we find no difference in the column
density distribution between blue and red galaxies ($1.9\sigma$, all
inclinations) along the major axis.

In Figure~\ref{fig:InclNvsV} we compare the cloud column densities and
velocities, and the pixel velocities for (a) face-on galaxies and (b)
edge-on galaxies, slicing the samples by galaxy color and azimuthal
angle. Similar to Figure~\ref{fig:PANvsV}, point colors and the left
and bottom histograms represent blue and red galaxy subsamples, while
point types and top and right histograms represent major axis and
minor axis subsamples. The indicated significances correspond to the
same statistical tests as in Figure~\ref{fig:PANvsV}. Again, the pixel
velocity F-test results in Figure~\ref{fig:InclNvsV} follow the TPCF
chi-squared test results, with the exception of the test comparing
edge-on galaxies probed along the major versus minor axes (panel
(b)). In this case the F-test result is significant, where edge-on,
major axis galaxies have a slightly larger dispersion in their pixel
velocities than for edge-on, minor axis galaxies. Due to the large
bootstrap uncertainties on the major axis TPCF, the chi-squared test
on the corresponding TPCFs is not significant (see
Figure~\ref{fig:PAincl}(e)).

We find that absorbers around face-on galaxies consist of clouds with
column densities that rarely exceed $\log N({\MgII}) = 14$. This is in
contrast to edge-on galaxies which are more likely to have these
higher column density clouds. For face-on galaxies, the column
densities of blue and red galaxies are statistically consistent
($0.3\sigma$). Comparing blue galaxies with face-on and edge-on
inclinations (for all azimuthal angles), we find that the column
densities of the edge-on subsample are larger than those in the
face-on subsample ($4.7\sigma$, not plotted). We also find that the
column densities of the blue, edge-on subsample are larger than the
red, edge-on sample ($4.0\sigma$). There are no such differences in
the column densities between galaxies probed along the major and minor
axes for face-on ($0.7\sigma$), edge-on ($0.9\sigma$), blue
($1.4\sigma$, not plotted) or red ($1.5\sigma$, not plotted) galaxies.

\section{Discussion}
\label{sec:discussion}

The results in the previous section demonstrate that absorber velocity
dispersions and cloud column densities of the gas traced by {\MgII}
absorption detected with $W_r(2796)\geq0.04$~{\AA} depend strongly on
where the gas is located around isolated galaxies. In particular, we
find that absorbers with the largest velocity spreads are associated
with blue galaxies that have ``face-on'' inclinations and are probed
along the galaxy projected minor axis. For ``edge-on'' galaxies, the
velocity structure is similar regardless of galaxy color or azimuthal
angle. Thus, the absorber velocity spreads (TPCFs) depend mostly on if
the galaxy is ``star-forming'' (using color as a proxy for the star
formation rate), or if the galaxy is ``face-on.'' The cloud column
densities depend mostly on galaxy inclination or color, where face-on
or red galaxies host absorbers with smaller cloud column densities
than those hosted by edge-on or blue galaxies.

Mechanisms in the baryon cycle which could give rise to these
differences in velocity structure and column densities of the
{\MgII}-absorbing gas include (1) merging satellite galaxies whose gas
is either being ejected due to star formation or being stripped by
tidal forces, (2) IGM accretion, recycled accretion, and/or rotating
material merging onto the galaxy disk, and (3) outflowing gas in
star-forming galaxies. In this section, we discuss which mechanisms
may most likely contribute to the differences in the observed velocity
dispersions and cloud column densities as a function of host galaxy
color (as a proxy for star formation) and orientation.

\subsection{Merging Satellite Galaxies}

Regardless of our results with orientation, previous works have
repeatedly shown that the probability of probing a satellite galaxy
with quasar absorption lines is low. \citet{martin12} found only one
satellite galaxy contributing to {\MgII} absorption out of a full
sample of over 200 galaxies, corresponding to a probability of finding
absorption due to a satellite galaxy of $<1\%$. \citet{tumlinson13}
used simulations to estimate the cross-section of {\HI} gas bound to
satellites and assumed that the internal velocity dispersion of gas in
the satellites cannot exceed the maximum circular velocity of the
satellite. They found that the mean number of satellites per sightline
is well below the number of absorption components per sightline at a
given velocity for their sample of {\HI} absorbers. For {\MgII}
absorption, \citet{gauthier10} examined covering fractions for
absorbers around luminous red galaxies (which are comparable to our
reddest galaxies) and compared them to the estimated maximum
cross-section of satellites in host halos. They found that this
maximum cross-section is much lower than the absorber covering
fractions at all impact parameters, and therefore satellites are not
sufficient for explaining the presence of the majority of absorption
around galaxies.

If we assume that the cross-section of satellites {\it is} great
enough to explain the presence of the majority of {\MgII} absorption,
we can investigate the preferential location of satellite galaxies
around a larger, central galaxy to determine if satellites could drive
our results. Many previous works have examined satellite distributions
and the consensus appears to be that, for red galaxies, satellites
within the virial radius are preferentially aligned with the central
galaxy's major axis, while there is no preferred alignment for
satellites around blue galaxies \citep[with the exception of galaxies
  in the Local Group;][and references therein]{yang06}. This has been
confirmed in recent simulations by \citet{dong14}. Our color cut
defining blue and red galaxies is slightly bluer than that used by
\citet{yang06} (our $B-K=1.399$ cut corresponds to $g-r=0.62$,
compared to their cut of $g-r=0.83$). Our bluer color cut may result
in more blue galaxies in our ``red'' sample than in the \citet{yang06}
red sample. Since the blue galaxies in Yang {\etal}~have been found to
have an isotropic distribution of satellites, we may have a somewhat
more isotropic distribution of satellites around red galaxies for our
sample.

We cannot rule out that we may be observing some amount of {\MgII}
absorption due to satellites. However, even though satellites align
with the major axis of red galaxies, we find no difference between the
TPCFs or column densities of our red galaxy samples along the major
and minor axes. The isotropic distribution of satellites around blue
galaxies is not consistent with the highly significant differences
between the TPCFs of our blue galaxy samples along the major and minor
axes. If satellites are the dominant source of {\MgII} absorption, we
would expect the TPCFs to show differences along the major and minor
axes in red galaxies (which could be mitigated by our bluer color
cut), but show no differences in blue galaxies (which we do find).

We do not have velocity information on the host galaxy--satellite
galaxy samples published in \citet{yang06}. However, the velocity
distribution of the satellites around galaxies is unlikely to be the
cause of the different absorber kinematic distributions for several
reasons. We are measuring the internal velocity dispersions of the
absorbers themselves, not the material with respect to the galaxy
systemic velocity. We assume our sightlines are unlikely to pass
through multiple satellites with a single sightline since the
probability of passing through a single galaxy is low ($<1\%$, from
above), and probing multiple satellites at once is even
lower. Therefore the internal gas velocity distributions of the
satellite galaxies would have to show differences for blue central
galaxies (to match the TPCFs), but not for red central
galaxies. However, if {\MgII} absorption is due to tidal stripping of
or outflows from the satellites, we would expect the satellites around
red galaxies (which tend to be located in more massive halos in our
sample) to be more disturbed and therefore have larger absorber
velocity dispersions than if they were located around blue galaxies,
which tend to be less massive in our sample. This is the reverse of
our findings. 

The galaxies in our sample were drawn from {\magiicat}
\citep{magiicat1}, which contains only galaxies that are isolated to
the limits of the data, where no spectroscopically identified neighbor
was found within $D<100$~kpc or a line-of-sight velocity of
500~{\kms}. If satellites are present in our sample, they are
undetected within both velocity space and impact
parameter. Additionally, all of the galaxies were imaged with {\it
  HST} and satellite galaxies were not detected in any of the images
to the limits of our data. Therefore, if we are probing satellite
galaxies, they are below our detection limits \citep[i.e., below
  roughly $0.1 L_B^{\ast}$, corresponding to a circular velocity of
  $\sim80$~{\kms} for these satellites;][]{magiicat1, cwc-masses2}.

\subsection{Accretion and Rotation}

Previous work examining the low-ionization CGM has found accreting
and/or rotating material for edge-on galaxies probed along the
projected major axis in the form of absorption that is shifted to one
side of the galaxy systemic velocity \citep[e.g.,][]{steidel02,
  ggk-sims, bouche13, rubin-accretion}. This is also commonly observed
in simulations \citep[e.g.,][]{stewart11}. Since we do not know
whether our absorbers are blueshifted or redshifted with respect to
the galaxy as can be determined through the down-the-barrel approach
used by \citet{rubin-accretion}, we cannot say for certain whether our
absorbers are accreting or rotating. However, we have examined the
properties of this material using absorber velocity dispersions and
cloud column densities for orientations in which accreting and
rotating material is expected to be found.

Edge-on galaxies probed along the major axis are most likely to
exhibit accreting or rotating material and are the least likely
orientation to be contaminated with outflowing material. The
line-of-sight velocity for rotation is greatest in edge-on galaxies
since the rotation is in the plane of the galaxy. In our sample, this
orientation is dominated by absorbers that contain clouds with larger
column densities ($\log N({\MgII}) > 13$), most of which have
velocities $v<|100|$~{\kms}. This is most easily seen as the triangle
points in Figure~\ref{fig:PANvsV}(b). We might expect to only see blue
galaxies with absorption in this orientation due to the red galaxies
having their star formation quenched from a lack of a gas reservoir,
but the subsample is populated by both blue and red galaxies. 

We could be seeing recycled accretion in galaxies of all colors from
past outflows \citep[e.g.,][]{oppenheimer10}. However, accretion onto
galaxies provides fuel for star formation, which we do not infer for
our red galaxies. It is possible that this material is rotating around
the galaxy and not accreting onto the galaxy, especially for the red
galaxies, though we cannot tell if this is the case for the data
presented here as our velocities are not shifted with respect to the
galaxy systemic velocity. Some mechanism may be suppressing star
formation in the red galaxies, but this mechanism is not present in
blue galaxies. It is also possible that some of the red galaxies we
have observed are in fact dusty, star-forming galaxies that are
currently accreting material, especially since our $B-K$ color cut
does include Sbc SED types in with the ``red'' galaxy subsample
\citep[for details concerning our $B-K$ color calculations,
  see][]{magiicat1}. This is less likely because we would have
observed signatures of outflows in both blue and red galaxies, unless
dusty, star-forming galaxies are rare in our sample. We do not yet
have the information needed to calculate the star formation rates to
determine if we have dusty, star-forming galaxies in our sample.

It is interesting that we find no significant difference in the TPCFs
of blue and red galaxies probed along the major axis (for all $i$) or
in edge-on inclinations (for all $\Phi$) in Figures~\ref{fig:BKPA}(a)
and \ref{fig:BKincl}(b), respectively. This suggests that the velocity
structure of absorbers along the major axis or for edge-on galaxies
does not depend on the star formation rate of the host galaxy, and
likely depends only on mass. This fits in with the accretion or
rotating gas picture, which, unlike outflows, should not depend on the
star formation rate unless the star formation rate is so great as to
prevent accretion and/or remove the gas reservoir. Also interesting is
that, for all $B-K$, the absorber velocity dispersions depend on
inclination for the major axis sample in
Figure~\ref{fig:PAincl}(c). Along the major axis, the absorber
velocity dispersion is greater for edge-on galaxies than for face-on
galaxies, and this may be due to rotating gas whose line-of-sight
velocity is maximized in edge-on inclinations, while the vertical
velocity dispersions in the disks of galaxies is small. The column
densities may also be greater for edge-on galaxies than for face-on
galaxies along the major axis (Figure~\ref{fig:PANvsV}(b)). This could
indicate that the accreting or rotating material for the edge-on
galaxies is more coherent, i.e., the path lengths or amount of
material are larger than for face-on galaxies.

\subsection{Outflows}

Absorption line studies have frequently found evidence for outflows as
blueshifted absorption in face-on galaxies
\citep[e.g.,][]{rubin-winds14} or as enhanced equivalent widths along
the projected minor axis for edge-on galaxies
\citep[e.g.,][]{bordoloi11, kcn12}. Therefore we also can associate
the kinematics results for our blue, face-on and minor axis subsamples
with outflows. Again, we cannot say for certain if our absorbers are
actually entrained in outflowing material so we instead focus on
possible absorption properties of the absorbers for orientations in
which outflows are most likely to occur.

For blue, edge-on galaxies probed along the minor axis (blue triangles
in Figure~\ref{fig:PANvsV}(a)), we found a group of lower column
density ($\log N({\MgII}) < 13$), higher velocity ($v \sim
|100|$~{\kms}) material. This is similar to \citet{fox15} who used a
background quasar whose line of sight passes through the biconical
Fermi Bubbles at the Milky Way Galactic center, where the Milky Way in
this case is similar to our edge-on, minor axis subsample. They
reported a complex absorption profile with two high-velocity metal
absorption components at $v_{\rm LSR} \sim \pm 250$~{\kms} with $\log
N({\SiII})\sim 13.2$. They associate these components with cool gas
that has been entrained in the near and far sides of an outflow from
the galactic center. In our sample, comparable components have similar
column densities, but a lower velocity with respect to the bulk of the
absorption, which may be due to the fact that we are probing outflows
at much larger impact parameters than \citet{fox15}; $D>20$~kpc
compared to $D=2.3$~kpc. This suggests that the low column density
clouds we find in our edge-on, minor axis galaxies may also be
associated with fragmented cool gas entrained in outflows.

In our subsample consisting of blue, face-on galaxies probed along the
minor axis (blue circles in Figure~\ref{fig:PANvsV}(a)), the absorbers
are dominated by low column density material ($\log N({\MgII})<14$,
where most of the clouds are $\log N({\MgII}) \leq 13$), with high
velocity components at $v>|100|$~{\kms}. This material is also similar
to the absorber found by \citet{fox15}. We suggest that the velocities
are larger for our face-on galaxies than for our edge-on galaxies
along the minor axis because we are observing down into the outflow
(i.e., close to the down-the-barrel approach) as opposed to across the
outflow. We expect the line-of-sight velocity dispersions to be larger
for the face-on sample as the biconical outflows are pointed toward
(or away from) the observer, and this is observed in our face-on, blue
galaxy subsample TPCFs.

The larger column density ($\log N({\MgII}) > 13$), lower velocity ($v
\lesssim |50|$~{\kms}) clouds for blue, edge-on galaxies probed along
the minor axis (seen in Figure~\ref{fig:PANvsV}(a) as blue triangles),
suggest that we are observing more material, a larger path length, a
larger metallicity, or ionization conditions that are more conducive
to {\MgII} absorption than for the face-on, minor axis galaxies (blue
circles) described in the previous paragraph. Since the material
probed in both of these orientations is most likely outflows, we do
not expect the metallicity or the ionization conditions to differ for
these two subsamples, and therefore should not be the cause of the
enhanced column densities for the edge-on sample compared to the
face-on sample. However, \citet{outflowsreview} listed large
variations in the metallicity and ionization levels between clouds as
a possible signature of outflows in quasar absorption lines. If this
were the case here, we should also see the large column density clouds
($\log N({\MgII})>13$) for face-on galaxies, but they are nonexistent
in our sample. Differences in the column densities with inclination
therefore hint at probing larger amounts of gas or larger path lengths
for the lowest velocity clouds associated with outflows in edge-on
galaxies (compared to face-on galaxies), an effect that may be due to
the geometry of the gas flow.

In contrast to the blue galaxies, we find very small absorber velocity
dispersions for red galaxies in face-on, minor axis orientations (red
circles in Figure~\ref{fig:PANvsV}(a)). We suggest that this is due to
a lack of current outflows. This is supported by the fact that redder
galaxies typically have lower star formation rates and do not
typically drive outflows. We caution that this subsample is small,
which may indicate that outflows in red galaxies for this orientation
are rare rather than nonexistent in general.

\section{Summary and Conclusions}
\label{sec:conclusions}

We examined the dependence of gas kinematics and cloud column
densities on galaxy rest-frame $B-K$ color, azimuthal angle, and
inclination for a subset of 30 isolated {\magiicat} \citep{magiicat1}
galaxies. Each galaxy was imaged with {\it HST} and was modeled with
GIM2D to obtain galaxy orientations. In each case, we have a
high-resolution optical spectrum of an association background quasar
within $20 < D < 110$~kpc for detailed kinematic analysis. To
characterize the absorber gas kinematics, we examined only those
absorbers detected with $W_r(2796)\geq0.04$~{\AA} and calculated
pixel-velocity two-point correlation functions (TPCFs) for galaxy
color, azimuthal angle, and inclination subsamples.

Our findings include the following:

\begin{enumerate}[noitemsep]

\item Absorption TPCFs with the longest high velocity separation tails
  (the largest velocity dispersions) are associated with blue
  galaxies, face-on galaxies, and galaxies probed along the minor
  axis. Conversely, the narrowest TPCF is associated with red, face-on
  galaxies. The velocity structure in the absorbers hosted by edge-on
  galaxies is similar regardless of galaxy color or azimuthal angle,
  and we find a lack of differences for red galaxies with azimuthal
  angle as well as for blue and red galaxies probed along the
  projected major axis.

\item Examining cloud column densities and velocities, we find, in
  general, that the highest velocity clouds with respect to the
  optical depth weighted mean of absorption ($z_{\rm abs}$) have small
  column densities, while the largest column density clouds are
  located at small velocities. When slicing our sample by galaxy color
  and inclination, we find that the column densities are larger for
  edge-on galaxies and blue galaxies than for face-on galaxies and red
  galaxies, respectively. The column densities show no dependence on
  azimuthal angle, however.

\item We find large absorber velocity dispersions and large column
  density clouds at low velocity for edge-on galaxies probed along the
  major axis, an orientation that is associated with gas accreting
  onto and/or rotating around galaxies. The velocity structure of the
  absorbers for this orientation does not depend on galaxy color and,
  by proxy, star formation rate. The large absorber velocity
  dispersions for this orientation may be due to rotating gas whose
  line-of-sight velocity is maximized due to the edge-on inclination,
  especially compared to the smaller velocity dispersions for face-on
  galaxies along the major axis. The large column density clouds may
  indicate that accreting and/or rotating material is fairly coherent,
  where the path lengths or amount of gas probed is larger than for
  other subsamples.

\item We associate the largest absorber velocity dispersions and
  smaller column density clouds at higher velocity for blue, face-on
  galaxies probed along the minor axis and for blue, edge-on galaxies
  along the minor axis with bipolar outflows, which are commonly
  observed in these orientations. The behavior of this material and
  the fact that it is located in regions which are associated with
  outflows may be an indication that the material probed by {\MgII}
  absorption is fragmented material entrained in outflows. These
  absorbers are similar to the absorption associated with the Fermi
  Bubbles in the Milky Way Galaxy \citep{fox15}. Larger column density
  clouds at low velocity are present for the blue, edge-on galaxies
  probed along the minor axis, and since comparable components are
  lacking for the face-on sample, we suggest that the path lengths of
  these structures are larger, possibly a geometry effect. For the
  smallest column density clouds ($\log N({\MgII}) \lesssim 13$), the
  cloud velocities (and absorber velocity dispersions) are greater for
  the face-on galaxies than edge-on galaxies. We attribute this to
  looking down into the outflow for face-on galaxies, and across the
  outflow for edge-on galaxies. Conversely, the very small absorber
  velocity dispersions for red galaxies at these orientations suggests
  that outflows are not active, a result of star formation having
  possibly been quenched.

\item Merging satellite galaxies may contribute to the observed
  {\MgII} absorption but are not expected to dominate the absorption
  due to their cross-section being too low. If they did dominate, then
  we would expect differences in the TPCFs to correspond to the
  preferred locations of the satellites around galaxies (isotropic
  distribution for blue galaxies and along the major axis for red
  galaxies), but this is not the case. We also do not expect the
  velocities of the satellites to have much of an effect on our TPCFs
  because it is unlikely that our sightlines are hitting multiple
  satellites at once, as the probability of detecting a single
  satellite is $\lesssim 1\%$ \citep{martin12}. Since red galaxies
  tend to be more massive than blue, we would expect larger internal
  gas velocity dispersions for satellites hosted by red central
  galaxies due to tidal stripping or outflows from the satellites
  themselves, but we do not find this to be the case.

\end{enumerate}

This work shows that the velocity structure and cloud column densities
of {\MgII} absorbers depend on a combination of galaxy orientations
and colors (and likely other galaxy properties not included in this
work), not just one or two properties at a time as has been commonly
examined in the past. For example, when observing edge-on galaxies,
absorption could be probing accretion along the major axis, or
outflows along the minor axis. In this case, the outflows will
dominate the absorption signature and the information for accretion
will likely be lost, which may contribute to the low covering
fractions found for accreting material.

We also find that while the equivalent width of an absorption profile
provides a useful diagnostic for generally determining what type of
material is being studied, considering both the velocity structure (in
terms of the velocity spread or dispersion) and cloud column densities
of absorbers with galaxy orientation is important for probing the
details of gas flows. This is evident in the fact that absorbers
probing outflows in face-on galaxies may have large equivalent widths
because of the large velocity spread in the absorbers, whereas
outflows in edge-on, minor axis galaxies may have large equivalent
widths mainly due to large column density clouds. Modeling the
ionization conditions and metallicities in the clouds would also be
beneficial, as large cloud-cloud variations in outflows have been
suggested by \citet{outflowsreview}, but this is beyond the scope of
the present work.

Future work might include examining mock quasar absorption line
spectra in simulations for a variety of galaxy colors (or star
formation rate and mass) and orientations to better understand the
origin of the absorber velocity structure and cloud column
densities. By applying our analysis to simulations, we not only obtain
a more physical explanation for what we observe, but we also help
constrain the CGM in simulations. Observationally, a larger sample of
galaxies with {\it HST} images and high-resolution spectra would allow
us to slice the sample by galaxy color, azimuthal angle, {\it and}
inclination to better explore the multivariate dependence of absorber
properties on the host galaxy and its baryon cycle. Our current sample
is only large enough to slice by two galaxy properties in the TPCFs,
though we have presented all three properties in
Figures~\ref{fig:PANvsV} and \ref{fig:InclNvsV}.

It would also be useful to obtain estimates of the absorber
metallicities to further understand whether we are probing cold-mode
accretion, recycled accretion, satellite material, and/or
outflows. Metallicities are important to help distinguish between
accretion and outflows, where accreting material is expected to have
lower metallicities than outflowing material \citep[see][and
  references therein]{lehner13}. Comparable metallicities across all
galaxy orientations might indicate that {\MgII} absorption is
primarily associated with outflows and recycled accretion, not
cold-mode accretion. Lastly, determining the star formation rates of
the galaxies in our sample is important to identify any possible
dusty, star forming galaxies which may be contaminating the results of
our red galaxy samples. The galaxy star formation rates provide a more
physical indicator of ongoing star formation than galaxy color.

\acknowledgments

We thank C.~Steidel for providing a reduced HIRES/Keck quasar
spectrum. This material is based upon work supported by the National
Science Foundation under Grant No. 1210200 (NSF East Asia and Pacific
Summer Institutes). N.M.N.~was also partially supported through a
NMSGC Graduate Fellowship and a Graduate Research Enhancement Grant
(GREG) sponsored by the Office of the Vice President for Research at
New Mexico State University. G.G.K.~acknowledges the support of the
Australian Research Council through the award of a Future Fellowship
(FT140100933). M.T.M.~thanks the Australian Research Council for
Discovery Project grant DP130100568 which supported this work.


\bibliographystyle{apj}
\bibliography{refs}

\end{document}